\documentclass[prd,twocolumn,showpacs,showkeys,floatfix]{revtex4}
\usepackage{amsmath,amssymb}
\usepackage{latexsym}
\usepackage{bm}

\usepackage{hyperref}
\usepackage[ansinew]{inputenc}
\makeatletter
\def\eqnarray{\stepcounter{equation}\let\@currentlabel=\theequation
\global\@eqnswtrue
\global\@eqcnt\z@\tabskip\@centering\let\\=\@eqncr
$$\halign to \displaywidth\bgroup\@eqnsel\hskip\@centering
  $\displaystyle\tabskip\z@{##}$&\global\@eqcnt\@ne
  \hfil${\;##\;}$\hfil
  &\global\@eqcnt\tw@ $\displaystyle\tabskip\z@{##}$\hfil
   \tabskip\@centering&\llap{##}\tabskip\z@\cr}
\makeatother

\begin{document}
\title{PHYSICAL CAUSES OF ENERGY-DENSITY INHOMOGENIZATION AND STABILITY OF ENERGY-DENSITY HOMOGENEITY IN  RELATIVISTIC SELF--GRAVITATING FLUIDS}
\author{L. Herrera\footnote{Also at U.C.V., Caracas}}
\email{laherrera@cantv.net.ve}
\affiliation{Departamento   de F\'{\i}sica Te\'orica e Historia de la  Ciencia,
Universidad del Pa\'{\i}s Vasco, Bilbao, Spain}

\begin{abstract}
We identify the factors responsible for  the appearance of energy--density inhomogeneities in a self--gravitating fluid, and describe  the evolution of those factors from an initially homogeneous distribution. It is shown that a specific combination of  the Weyl tensor and/or local anisotropy of pressure and/or  dissipative fluxes entails the formation of energy--density inhomogeneities. Different cases are analyzed in detail and  in the particular case of dissipative fluids, the role of relaxational processes  as well as non--local effects are brought out.
\end{abstract}

\date{\today}
\maketitle
\section{INTRODUCTION}
The role of energy--density inhomogeneity in the collapse of self--gravitating fluids  is quite relevant and has been extensively discussed in the literature
(see \cite{1}--\cite{11chan} and references therein).

Furthermore the Penrose's proposal \cite{11} to define a gravitational arrow of time in terms of the Weyl tensor,  is based on the particularly  simple relation between the Weyl tensor and
density inhomogeneity, for perfect fluids. 
However the fact that such a relationship is no longer valid in the presence
of local anisotropy of the pressure and/or
dissipative processes, and/or electric charge distribution \cite{12, 13, 14bis}, explains its
failure in scenarios where the above-mentioned  factors are present \cite{14, 15, 16}.
Therefore, since the rationale
behind Penrose's idea is that tidal forces tend to make the
gravitating fluid more inhomogeneous as the evolution proceeds,
thereby indicating the sense of time, it should be clear that all factors asociated to energy--density inhomogenity (and not only the Weyl tensor)  should be present in any definition of the gravitational arrow of time {\it a la} Penrose.

However, in spite of its relevance, it is unclear yet how different physical phenomena affect and, more precisely, produce energy-  density inhomogeneities. The pertinence of such a question is supported by the arguments above.

It is the purpose of this work to answer to the above mentioned question. It  entails in fact three  different (but related)  questions, namely:
\begin{itemize}
\item What   aspects of the fluid distribution are related  (and how) to the existence of energy--density inhomogeneities ?
\item How those factors evolve, starting with an initially homogeneous distribution? 
\end{itemize}
this last question in turn leads to a third relevant question:
\begin{itemize}
\item Under which conditions an initially homogeneous configuration remains homogeneous all along its evolution (stability problem)?
\end{itemize}
Our procedure will be heavily rely on two differential  equations  relating the Weyl tensor to different physical variables. On of them  is an evolution equation containing time derivatives of  those variables, whereas the other is a constraint equation implying spatial derivatives. As far as we know these equations were first derived by Ellis \cite{ellis1, ellis2} for configurations without any specific symmetry,  afterwards they have been reobtained and used by different authors (see for example \cite{12}, \cite{ellis4}--\cite{she}).

 We shall consider general fluid distributions endowed with anisotropic pressure and dissipating energy during its evolution. The specific physical (microscopic) phenomena behind these fluid characteristics will be not discussed here, instead we shall be concerned only by the macroscopic (hydrodynamic) manifestations of those phenomena  (arguments to justify such kind of fluid distributions may be found in \cite{25}--\cite{27} ). 
 
  Dissipation processes are usually treated invoking  two possible (opposite) approximations: diffusion and streaming out, so we shall do here. In the case of diffusion approximation a causal transport equation will be used, allowing to bring out the effects of pre--relaxational phenomena.

In the specific case of localized  configurations we have to assume that our fluid distribution is bounded by a spherical surface. In order to avoid thin shells on such a boundary surface Darmois \cite{28} conditions should be imposed.

\section{ENERGY--MOMENTUM TENSOR,  RELEVANT VARIABLES AND FIELD EQUATIONS}
As is usually assumed in the study of self--gravitating compact objects, we shall consider that deviations from spherical symmetry are  incidental rather than basic features of the process involved. Accordingly we shall restrain ourselves to spherically symmetric fluid distributions. 

Thus, we consider a spherically symmetric distribution  of collapsing
fluid, bounded by a spherical surface $\Sigma$. The fluid is
assumed to be locally anisotropic (principal stresses unequal) and undergoing dissipation in the
form of heat flow (to model dissipation in the diffusion approximation) and  null radiation (to model dissipation in the free streaming approximation).

Choosing comoving coordinates inside $\Sigma$, the general
interior metric can be written
\begin{equation}
ds^2=-A^2dt^2+B^2dr^2+R^2(d\theta^2+\sin^2\theta d\phi^2),
\label{1m}
\end{equation}
where $A$, $B$ and $R$ are functions of $t$ and $r$ and are assumed
positive. We number the coordinates $x^0=t$, $x^1=r$, $x^2=\theta$
and $x^3=\phi$.

The matter energy-momentum $T_{\alpha\beta}$ inside $\Sigma$
has the form
\begin{eqnarray}
T_{\alpha\beta}&=&(\mu +
P_{\perp})V_{\alpha}V_{\beta}+P_{\perp}g_{\alpha\beta}+(P_r-P_{\perp})\chi_{
\alpha}\chi_{\beta}+q_{\alpha}V_{\beta}\nonumber \\&+&V_{\alpha}q_{\beta}+
\epsilon l_{\alpha}l_{\beta}, \label{3}
\end{eqnarray}
where $\mu$ is the energy density, $P_r$ the radial pressure,
$P_{\perp}$ the tangential pressure, $q^{\alpha}$ the heat flux describing dissipation in the diffusion approximation,
$\epsilon$ the energy density of the null fluid describing dissipation in the free streaming approximation, $V^{\alpha}$ the four velocity of the fluid,
$\chi^{\alpha}$ a unit four vector along the radial direction
and $l^{\alpha}$ a radial null four vector. These quantities
satisfy
\begin{equation}
V^{\alpha}V_{\alpha}=-1, \;\; V^{\alpha}q_{\alpha}=0, \;\; \chi^{\alpha}\chi_{\alpha}=1, 
\end{equation}
and 
\begin{equation}
\chi^{\alpha}V_{\alpha}=0, \;\; l^{\alpha}V_{\alpha}=-1, \;\; l^{\alpha}l_{\alpha}=0.
\end{equation}

The four--acceleration $a_{\alpha}$ and the expansion $\Theta$ of the fluid are
given by
\begin{equation}
a_{\alpha}=V_{\alpha ;\beta}V^{\beta}, \;\;
\Theta={V^{\alpha}}_{;\alpha}. \label{4b}
\end{equation}
and its  shear $\sigma_{\alpha\beta}$ by
\begin{equation}
\sigma_{\alpha\beta}=V_{(\alpha
;\beta)}+a_{(\alpha}V_{\beta)}-\frac{1}{3}\Theta h_{\alpha \beta},\label{4a}
\end{equation}
where $h_{\alpha \beta}=g_{\alpha\beta}+V_{\alpha}V
_{\beta}
.$

We do not explicitly add bulk viscosity and/or shear viscosity to the system because they
can be absorbed into the radial and tangential pressures, $P_r$ and
$P_{\perp}$, of the
collapsing fluid.

Since we assumed the metric (\ref{1m}) comoving then
\begin{eqnarray}
V^{\alpha}=A^{-1}\delta_0^{\alpha}, \;\;
q^{\alpha}=qB^{-1}\delta^{\alpha}_1, \;\;\nonumber \\
l^{\alpha}=A^{-1}\delta^{\alpha}_0+B^{-1}\delta^{\alpha}_1, \;\;
\chi^{\alpha}=B^{-1}\delta^{\alpha}_1, \label{5}
\end{eqnarray}
where $q$ is a function of $t$ and $r$ satisfying $q^\alpha = q \chi^\alpha$.

From  (\ref{4b}) with (\ref{5}) we have for the  acceleration and its scalar $a$,
\begin{equation}
a_1=\frac{A^{\prime}}{A}, \;\; a^2=a^{\alpha}a_{\alpha}=\left(\frac{A^{\prime}}{AB}\right)^2, \label{5c}
\end{equation}
where $a^\alpha= a \chi^\alpha$,
and for the expansion
\begin{equation}
\Theta=\frac{1}{A}\left(\frac{\dot{B}}{B}+2\frac{\dot{R}}{R}\right),
\label{5c1}
\end{equation}
where the  prime stands for $r$
differentiation and the dot stands for differentiation with respect to $t$.
With (\ref{5}) we obtain
for the shear (\ref{4a}) its non zero components
\begin{equation}
\sigma_{11}=\frac{2}{3}B^2\sigma, \;\;
\sigma_{22}=\frac{\sigma_{33}}{\sin^2\theta}=-\frac{1}{3}R^2\sigma,
 \label{5a}
\end{equation}
and its scalar
\begin{equation}
\sigma^{\alpha\beta}\sigma_{\alpha\beta}=\frac{2}{3}\sigma^2,
\label{5b}
\end{equation}
where
\begin{equation}
\sigma=\frac{1}{A}\left(\frac{\dot{B}}{B}-\frac{\dot{R}}{R}\right).\label{5b1}
\end{equation}
Then, the shear tensor can be written as
\begin{equation}
\sigma_{\alpha \beta}= \sigma \left(\chi_\alpha \chi_\beta - \frac{1}{3} h_{\alpha \beta}\right).
\label{sh}
\end{equation}

\subsection{The Einstein equations}

Einstein's field equations for the metric (\ref{1m}) are given by
\begin{equation}
G_{\alpha \beta} = 8 \pi T_{\alpha \beta},
\label{Eeq}
\end{equation}
 its non zero components
with (\ref{1m}), (\ref{3}) and (\ref{5}) become
\begin{widetext}
\begin{eqnarray}
8\pi T_{00}=8\pi(\mu+\epsilon)A^2
=\left(2\frac{\dot{B}}{B}+\frac{\dot{R}}{R}\right)\frac{\dot{R}}{R}
-\left(\frac{A}{B}\right)^2\left[2\frac{R^{\prime\prime}}{R}+\left(\frac{R^{\prime}}{R}\right)^2
-2\frac{B^{\prime}}{B}\frac{R^{\prime}}{R}-\left(\frac{B}{R}\right)^2\right],
\label{12} 
\end{eqnarray}
\end{widetext}
\begin{widetext}
\begin{eqnarray}
8\pi T_{01}=-8\pi(q+\epsilon)AB
=-2\left(\frac{{\dot R}^{\prime}}{R}
-\frac{\dot B}{B}\frac{R^{\prime}}{R}-\frac{\dot
R}{R}\frac{A^{\prime}}{A}\right),
\label{13} 
\end{eqnarray}
\end{widetext}
\begin{widetext}
\begin{eqnarray}
8\pi T_{11}=8\pi (P_r +\epsilon) B^2
=-\left(\frac{B}{A}\right)^2\left[2\frac{\ddot{R}}{R}-\left(2\frac{\dot A}{A}-\frac{\dot{R}}{R}\right)
\frac{\dot R}{R}\right]
+\left(2\frac{A^{\prime}}{A}+\frac{R^{\prime}}{R}\right)\frac{R^{\prime}}{R}-\left(\frac{B}{R}\right)^2,
\label{14} 
\end{eqnarray}
\end{widetext}
\begin{widetext}
\begin{eqnarray}
8\pi T_{22}=\frac{8\pi}{\sin^2\theta}T_{33}=8\pi P_{\perp}R^2
=-\left(\frac{R}{A}\right)^2\left[\frac{\ddot{B}}{B}+\frac{\ddot{R}}{R}
-\frac{\dot{A}}{A}\left(\frac{\dot{B}}{B}+\frac{\dot{R}}{R}\right)
+\frac{\dot{B}}{B}\frac{\dot{R}}{R}\right]\nonumber \\
+\left(\frac{R}{B}\right)^2\left[\frac{A^{\prime\prime}}{A}
+\frac{R^{\prime\prime}}{R}-\frac{A^{\prime}}{A}\frac{B^{\prime}}{B}
+\left(\frac{A^{\prime}}{A}-\frac{B^{\prime}}{B}\right)\frac{R^{\prime}}{R}\right].\label{15}
\end{eqnarray}
\end{widetext}
\subsection{The mass function}
Let us now introduce the mass function $m(t,r)$ \cite{Misner} (see also \cite{Cahill}), defined by
\begin{equation}
m=\frac{R^3}{2}{R_{23}}^{23}
=\frac{R}{2}\left[\left(\frac{\dot R}{A}\right)^2-\left(\frac{R^{\prime}}{B}\right)^2+1\right].
 \label{17masa}
\end{equation}

Following Misner and Sharp \cite{Misner}, it  is useful to define the proper time derivative $D_T$
given by
\begin{equation}
D_T=\frac{1}{A}\frac{\partial}{\partial t}, \label{16}
\end{equation}
and the``$R$'' derivative $D_R$,
\begin{equation}
D_R=\frac{1}{R^{\prime}}\frac{\partial}{\partial r}, \label{23a}
\end{equation}
where $R$ defines the areal radius of a spherical surface inside $\Sigma$ ( as
measured from its area).

Using (\ref{16}) we can define the velocity $U$ of the collapsing
fluid  as the variation of the areal radius with respect to proper time, i.e.
\begin{equation}
U=D_TR. \label{19}
\end{equation}
Then (\ref{17masa}) can be rewritten as
\begin{equation}
E \equiv \frac{R^{\prime}}{B}=\left(1+U^2-\frac{2m}{R}\right)^{1/2}.
\label{20x}
\end{equation}

Using (\ref{12})-(\ref{14}) with (\ref{16}) and (\ref{23a}) we obtain from
(\ref{17masa})
\begin{eqnarray}
D_Tm=-4\pi\left[\left
(\tilde{P}_r-\frac{4}{3}\eta\sigma\right)U+\tilde{q}E\right]R^2,
\label{22Dt}
\end{eqnarray}
and
\begin{eqnarray}
D_Rm=4\pi\left(\tilde{\mu}+\tilde{q}\frac{U}{E}\right)R^2,
\label{27Dr}
\end{eqnarray}
which implies
\begin{equation}
m=4\pi\int^{r}_{0}\left(\tilde{\mu}
+\tilde{q}\frac{U}{E}\right)R^2R'dr, \label{27intcopy}
\end{equation}
(assuming a regular centre to the distribution, so $m(0)=0$).
Integrating (\ref{27intcopy}) we find
\begin{equation}
\frac{3m}{R^3} = 4\pi\tilde{\mu} - \frac{4\pi}{R^3} \int^r_0{R^3\left(D_R{\tilde \mu}-3 \tilde q \frac{U}{RE}\right) R'dr}.
\label{3m/R3}
\end{equation}
with 
$$\tilde \mu= \mu+\epsilon,$$
$$\tilde P_{r}=P_r+\epsilon,$$
$$\tilde q= q+\epsilon,$$
\subsection{ Weyl tensor}

The Weyl tensor is defined through the  Riemann tensor
$R^{\rho}_{\alpha \beta \mu}$, the  Ricci tensor
$R_{\alpha\beta}$ and the curvature scalar $\cal R$, as:
$$
C^{\rho}_{\alpha \beta \mu}=R^\rho_{\alpha \beta \mu}-\frac{1}{2}
R^\rho_{\beta}g_{\alpha \mu}+\frac{1}{2}R_{\alpha \beta}\delta
^\rho_{\mu}-\frac{1}{2}R_{\alpha \mu}\delta^\rho_\beta$$
\begin{equation}
+\frac{1}{2}R^\rho_\mu g_{\alpha \beta}+\frac{1}{6}{\cal
R}(\delta^\rho_\beta g_{\alpha \mu}-g_{\alpha
\beta}\delta^\rho_\mu). \label{34}
\end{equation}

The electric  part of  Weyl tensor is defined by
\begin{equation}
E_{\alpha \beta} = C_{\alpha \mu \beta \nu} V^\mu V^\nu,
\label{elec}
\end{equation}
with the following non--vanishing components
\begin{eqnarray}
E_{11}&=&\frac{2}{3}B^2 {\cal E},\nonumber \\
E_{22}&=&-\frac{1}{3} R^2 {\cal E}, \nonumber \\
E_{33}&=& E_{22} \sin^2{\theta}, \label{ecomp}
\end{eqnarray}
where
\begin{eqnarray}
&&{\cal E}= \frac{1}{2}\left[\frac{\ddot R}{R} - \frac{\ddot B}{B} - \left(\frac{\dot R}{R} - \frac{\dot B}{B}\right) \frac{\dot R}{R}\right]\nonumber \\
&+& \frac{1}{2 B^2} \left[ -
\frac{R^{\prime\prime}}{R} + \left(\frac{B^{\prime}}{B} +
\frac{R^{\prime}}{R}\right) \frac{R^{\prime}}{R}\right]
-\frac{1}{2 R^2}. \label{E}
\end{eqnarray}

Observe that we may also write $E_{\alpha\beta}$ as:
\begin{equation}
E_{\alpha \beta}={\cal E} (\chi_\alpha
\chi_\beta-\frac{1}{3}h_{\alpha \beta}). \label{52}
\end{equation}

Finally, using (\ref{12}), (\ref{14}), (\ref{15}) with (\ref{17masa}) and (\ref{E}) we obtain
\begin{equation}
\frac{3m}{R^3} = 4 \pi (\tilde \mu - \Pi ) - {\cal E},
\label{8}
\end{equation}
with 
$$\Pi=\tilde P_{r}-P_{\bot}.$$

\section{ BIANCHI IDENTITIES AND ELLIS EQUATIONS }

As mentioned in the Introduction two differential equations for the Weyl tensor will play a central role in our work, these two equations originally found by Ellis \cite{ellis1, ellis2} are here reobtained following the procedure adopted in \cite{12}. Beside these two equations we shall also need the Bianchi identities, which for the system under consideration have  two independent components  which read (see \cite{H1} for details):
\begin{widetext}
\begin{equation}
\dot{\tilde \mu}+\left(\tilde \mu + \tilde P_r \right)\frac{\dot B}{B}+2\left(\tilde \mu + P_\bot \right)\frac{\dot R}{R}+ \frac{\tilde q^\prime A}{B} + 2 \tilde q \frac{(AR)^\prime}{BR}=0,
\label{1}
\end{equation}
\end{widetext}
and
\begin{widetext}
\begin{equation}
\dot{\tilde q} +\left(\tilde P_r\right)^\prime  \frac{A}{B}+ 2 \tilde q \left(\frac{\dot B}{B}+\frac{\dot R}{R} \right) + \left(\tilde \mu + \tilde P_r \right)\frac{A^\prime}{B} + 2 \Pi \frac{A R^\prime }{B R}=0.
\label{2}
\end{equation}
\end{widetext}

Finally, the following  equations for the Weyl tensor may be derived (e.g. see \cite{12} for details, but notice changes in notation)
\begin{widetext}
\begin{eqnarray}
\left[{\cal E }- 4\pi\left(\tilde \mu - \Pi \right)\right]^{\dot{}}&=& \frac{3\dot R}{R}\left(\frac{3m}{R^3} + 4\pi \tilde P_r \right)+ 12 \pi \tilde q \frac{A R^\prime}{B R}
= \frac{3\dot R}{R}\left[4 \pi \left(\tilde \mu + P_\bot \right)-{\cal E}\right] + 12 \pi \tilde q \frac{A R^\prime}{B R},
\label{6}
\end{eqnarray}
\end{widetext}
\begin{widetext}
\begin{eqnarray}
\left[{\cal E} - 4\pi\left(\tilde \mu - \Pi \right)\right]^\prime=-\frac{3 R^\prime}{R}\left({\cal E} + 4 \pi \Pi \right) - 12 \pi \tilde q \frac{B \dot R}{A R}
=-\frac{3 R^\prime}{R}\left[4 \pi \tilde \mu - \frac{3m}{R^3}\right] - 12 \pi \tilde q \frac{B \dot R}{A R}.
\label{7}
\end{eqnarray}
\end{widetext}

\section{THE TRANSPORT EQUATION}
 In the diffusion approximation ($\epsilon=0, \tilde q=q$), we shall need a transport equation derived from  a causal  dissipative theory ( e.g. the
M\"{u}ller-Israel-Stewart second
order phenomenological theory for dissipative fluids \cite{Muller67, IsSt76, I, II}).

Indeed, as it is already  well known the Maxwell-Fourier law for
radiation flux leads to a parabolic equation (diffusion equation)
which predicts propagation of perturbations with infinite speed
(see \cite{6D}-\cite{8'} and references therein). This simple fact
is at the origin of the pathologies \cite{9H} found in the
approaches of Eckart \cite{10E} and Landau \cite{11L} for
relativistic dissipative processes. To overcome such difficulties,
various relativistic
theories with non-vanishing relaxation times have been proposed in
the past \cite{Muller67,IsSt76, I, II, 14Di,15d}. Although the final word on this issue has not yet been said, the important point is that
all these theories provide a heat transport equation which is not
of Maxwell-Fourier type but of Cattaneo type \cite{18D}, leading
thereby to a hyperbolic equation for the propagation of thermal
perturbations.

A key quantity in these theories is the relaxation time $\tau$ of the
corresponding  dissipative process. This positive--definite quantity has a
distinct physical meaning, namely the time taken by the system to return
spontaneously to the steady state (whether of thermodynamic equilibrium or
not) after it has been suddenly removed from it. 
Therefore, when studying transient regimes, i.e., the evolution from a 
steady--state situation to a new one,  $\tau$ cannot be neglected. In 
fact, leaving aside that parabolic theories are necessarily non--causal,
it is obvious that whenever the time scale of the problem under
consideration becomes of the order of (or smaller) than the relaxation time,
the latter cannot be ignored. Indeed, 
neglecting the relaxation time ammounts -in this situation- to
disregarding the whole problem under consideration.

It is worth mentioning that sometimes in the past it has been argued that dissipative processes with relaxation times
comparable  to the characteristic time of the system are out of the
hydrodynamic regime.  However, the concept of hydrodynamic regime
involves the ratio between the  mean free path of fluid particles  and
the characteristic length of the system. When this ratio is lower that
unity, the fluid is within the hydrodynamic regime. When it is larger
than unity, the regime becomes Knudsen's. In the latter case the fluid is
no longer a continuum and even hyperbolic theories cease to be realiable.

Therefore that argument  can be valid only if the particles making up the
fluid are the same ones that transport the heat. However, this is 
never the case. Specifically, for a neutron star, $\tau$ is of the
order of the scattering time between electrons (which carry the
heat) but this fact is not an obstacle (no matter how large the
mean free path of these electrons may be) to consider the neutron
star as formed by a Fermi fluid of degenerate neutrons. The same
is true for the second sound in superfluid Helium and solids, and
for almost any ordinary fluid. In brief, the hydrodynamic regime
refers to fluid particles that not necessarily (and as a matter of fact,
almost never) transport the heat. Therefore large relaxation times (large
mean free paths of particles involved in heat transport) does not imply a
departure from the hydrodynamic regime (this fact has been streseed before
\cite{Santos}, but it is usually overlooked).

The corresponding  transport equation for the heat flux reads
\begin{equation}
\tau
h^{\alpha\beta}V^{\gamma}q_{\beta;\gamma}+q^{\alpha}=-\kappa h^{\alpha\beta}
(T_{,\beta}+Ta_{\beta}) -\frac 12\kappa T^2\left( \frac{\tau
V^\beta }{\kappa T^2}\right) _{;\beta }q^\alpha ,  \label{21t}
\end{equation}
where $\kappa $  denotes the thermal conductivity, and  $T$ and
$\tau$ denote temperature and relaxation time respectively. Observe
that, due to the symmetry of the problem, equation (\ref{21t}) only
has one independent component, which may be written as
\begin{equation}
\tau{\dot q}=-\frac{1}{2}\kappa qT^2\left(\frac{\tau}{\kappa
T^2}\right)^{\dot{}}-\frac{1}{2}\tau q\Theta A-\frac{\kappa}{B}(TA)^{\prime}-qA.
\label{te}
\end{equation}
In the case $\tau=0$ we recover the Eckart--Landau equation.

In the truncated version of the theory, the last term in (\ref{21t}) is absent (see for example \cite{pavon}), and (\ref{te}) becomes
\begin{equation}
\tau \dot{q}+qA=-\frac{\kappa}{B} (TA)^{\prime}.
\label{ISE'bis}
\end{equation}
\section{CAUSES OF INHOMOGENIZATION AND STABILITY OF  HOMOGENEOUS ENERGY--DENSITY CONDITION}
We shall now use the two equations for the Weyl tensor introduced in section III in order to answer to the main questions raised in this work. We shall proceed by considering particular cases, in an increasing order of complexity

\subsection{Non--dissipative dust} 
In this case $P_r=P_{\perp}=\tilde q=0$ and from the fact that the fluid is geodesic as follows from (\ref{2}) we can put  $A=1$.

Then equations (\ref{6}) and (\ref{7}) become
\begin{eqnarray}
\left({\cal E} - 4\pi \mu \right)^{\dot{}}+\frac{3\dot R}{R}\left({\cal E}-4 \pi \mu \right) =0,
\label{6dnd}
\end{eqnarray}
and 
\begin{eqnarray}
\left({\cal E}- 4\pi \mu \right)^\prime=-\frac{3 R^\prime}{R}{\cal E }.
\label{7dnd}
\end{eqnarray}
From this last equation it follows at once that ${\cal E}=0 \leftrightarrow \mu'=0$ a well known result implying that energy--density inhomogeneities in this particular case are controlled by the Weyl tensor alone.

Next, using (\ref{5b1}) and (\ref{1}) in (\ref{6dnd}), this last equation takes the form
\begin{equation}
\dot {\cal E}+\frac{3{\cal E}\dot R}{R}=-4\pi \mu \sigma,
\label{ndd1}
\end{equation}
whose solution may be written as
\begin{equation}
{\cal E}=-\frac{4 \pi \int^t_0 \mu \sigma R^3 dt}{R^3},
\label{ndd2}
\end{equation}
where we have chosen the integration function such that ${\cal E}(0,r)=0$.

It is worth noticing that in this particular case (non--dissipative dust), conformal flatness and shear--free conditions imply each other as it follows from (\ref{ndd1}) and the evolution equation for the shear, which reads (see \cite{she, LTB})
\begin{equation}
\dot \sigma+\frac{\sigma^2}{3}+\frac{2 \Theta \sigma}{3}=-{\cal E}.
\label{ndd3}
\end{equation}

Thus, an initially homogeneous distribution of non--dissipative dust will remain so iff it  remains conformally flat, which in turn implies that it remains shear--free. In other words the problem of  the stability of the energy--density homogenity reduces to the stability of conformal flatness or to the stability of the shear--free condition.
\subsection{Non--dissipative locally isotropic fluid}
The next case in increasing order of complexity, corresponds to a non--dissipative isotropic (in the pressure) fluid.  Thus we have $\Pi=\tilde q=0, P_r=P_{\perp}=P$. Then equations (\ref{6}) and (\ref{7}) become
\begin{eqnarray}
\left({\cal E} - 4\pi \mu \right)^{\dot{}}+\frac{3\dot R}{R}\left[{\cal E}-4 \pi (\mu +P)\right] =0,
\label{6disond}
\end{eqnarray}
and 
\begin{eqnarray}
\left({\cal E}- 4\pi \mu \right)^\prime=-\frac{3 R^\prime}{R}{\cal E }.
\label{7isond}
\end{eqnarray}
As in the previous case we have  from (\ref{7isond}) that ${\cal E}=0 \leftrightarrow \mu'=0$, meaning that 
energy--density inhomogeneities are controlled by the Weyl tensor alone, in this case too. In other words, the problem of the stability of the energy--density homogeneity reduces to that of the stability of the conformal flatness. It is also worth mentioning that in this case the vanishing of Weyl implies shear--free, however the inverse  is no true, since the evolution equation for the shear now reads
\begin{equation}
{\cal E} = \frac{a^\prime}{B}-\frac{\dot \sigma}{A}+a^2-\frac{\sigma^2}{3}-\frac{2}{3}\Theta \sigma-a\frac{R^\prime}{BR}.
\label{5es}
\end{equation}

Next, using using (\ref{5b1}) and (\ref{1}) in (\ref{6disond}), this last equation becomes
\begin{equation}
\dot {\cal E}+\frac{3{\cal E}\dot R}{R}=-4\pi (\mu+P) A\sigma.
\label{ndisod1}
\end{equation}

In the general shearing case we have from (\ref{ndisod1}) 
\begin{equation}
{\cal E}=-\frac{4 \pi \int^t_0 (\mu +P)A\sigma R^3 dt}{R^3},
\label{nddios2}
\end{equation}
where again we have chosen the integration function such that ${\cal E}(0,r)=0$. Obviously, in the case of dust (\ref{nddios2}) becomes (\ref{ndd2}).

Thus the stability of the energy--density homogeneity condition is as in the previos case controlled by the shear  of the fluid.

However, as mentioned before,  even though all spherically symmetric
conformally flat spacetimes (without dissipation) and isotropic fluids are shear-free the inverse is not true, and we may assume the fluid to be shear--free without that  implying conformal flatness. In this latter case we have from (\ref{ndisod1})
\begin{equation}
{\cal E}=\frac{f(r)}{R^3},
\label{ni1}
\end{equation}
where $f(r)$ is an arbitrary function of $r$ satisfying $f(0)=0$.

Then, if we start with an initially  homogeneous configuration at $t=0$, implying ${\cal E}(0,r)=0$ then $f(r)=0$ producing ${\cal E}=0$ for any $t$ and accordingly the energy--density homogeneous condition will hold all along the evolution.

If instead we assume a small though non--vanishing initial Weyl tensor then it will remain small if the fluid expands. However in the contraction case it may become relevant for sufficiently small $R$. Thus, in the case of contraction the energy--density homogeneity  might not be stable in the Lyapounov sense for a  shear--free perfect fluid.
\subsection{Locally anisotropic non--dissipative fluid}
We shall now bring out the role of the pressure anisotropy in the problem under consideration. For that purpose we shall assume $\tilde q=0$ but $\Pi\neq 0$. Then  equations (\ref{6}) and (\ref{7}) become
\begin{eqnarray}
\left({\cal E} - 4\pi \mu +4\pi \Pi\right)^{\dot{}}+\frac{3\dot R}{R}\left[{\cal E}-4 \pi (\mu +P_{\perp})\right] =0,
\label{6diansond}
\end{eqnarray}
and 
\begin{eqnarray}
\left({\cal E} - 4\pi \mu +4\pi \Pi\right)^{\prime{}}+\frac{3 R'}{R}\left({\cal E}+4 \pi \Pi\right) =0.
\label{7diansond}
\end{eqnarray}

The first remark emerging from (\ref{7diansond})  is that unlike the previous cases, now the responsible for the existence of energy--density inhomogeneity is not the Weyl tensor alone, but the quantity ${\cal E}+4\pi \Pi$. Indeed, assuming that the fluid distribution is regular at the centre,  it follows at once from (\ref{7diansond}) that ${\cal E}+4\pi \Pi=0 \leftrightarrow \mu'=0$.

Next, using (\ref{5b1}) and (\ref{1}) in (\ref{6diansond}), this last equation becomes
\begin{equation}
({\cal E}+4\pi \Pi)^{\dot{}}+ \frac{(3{\cal E}+4\pi \Pi)\dot R}{R}=-4\pi (\mu+P_r) A\sigma,
\label{ndanisod1}
\end{equation}
thus we have an evolution equation for the quantity responsible for the energy--density inhomogenity.
It is worth observing that the quantity  $-({\cal E}+4\pi \Pi)$ is referred to in previous works  \cite{sp}, \cite{LTB} (notice a difference in notation with respect to the first of these references) as  one of the structure scalars ($ X_{TF}$). It is defined as follows:

Let us introduce  the tensor
$X_{\alpha \beta}$ defined
by:

\begin{equation}
X_{\alpha \beta}=^*R^{*}_{\alpha \gamma \beta \delta}V^\gamma
V^\delta=\frac{1}{2}\eta_{\alpha\gamma}^{\quad \epsilon
\rho}R^{*}_{\epsilon \rho\beta\delta}V^\gamma V^\delta,
\label{magnetic}
\end{equation}

\noindent where $R^*_{\alpha \beta \gamma \delta}=\frac{1}{2}\eta
_{\epsilon \rho \gamma \delta} R_{\alpha \beta}^{\quad \epsilon
\rho}$ and $\eta
_{\epsilon \rho \gamma \delta}$ denotes the Levi--Civita tensor.

\noindent Tensor  $X_{\alpha \beta}$ may also be expressed through its trace and its trace--free part, as
\begin{eqnarray}
X_{\alpha\beta}=\frac{1}{3}X_T h_{\alpha
\beta}+X_{TF}(\chi_{\alpha} \chi_{\beta}-\frac{1}{3}h_{\alpha
\beta}).\label{magnetic'}
\end{eqnarray}

Then from  field equations, (\ref{magnetic'}) and using (\ref{34})
 (\ref{E})  and  (\ref{52}), we obtain (assuming vanishing dissipation)

\begin{equation}
X_{TF}=-{\cal E}-4\pi \Pi \label{EX}.
\end{equation}

Therefore the evolution equation (\ref{ndanisod1}) in terms of $X_{TF}$ reads
\begin{equation}
\dot X_{TF}+ \frac{3 X_{TF} \dot R}{R}=-8\pi \Pi \frac{\dot R}{R}+4\pi (\mu+P_r) A\sigma,
\label{ndanisod1b}
\end{equation}
whose general solution is
\begin{equation}
X_{TF}=-\frac{4 \pi \int^t_0 \left[2\Pi \dot R-(\mu +P_r)AR\sigma \right]R^2 dt}{R^3}.
\label{anddios2}
\end{equation}
Thus our system is initially homogeneous in the energy--density, however as time goes on this condition will be affected by the anisotropy of the pressure and the shear of the fluid according to (\ref{anddios2}).
\subsection{Dissipative  geodesic dust.}
Finally, in order to bring out the effect of dissipation in the formation of energy--density inhomogeneities we shall consider the case of dissipative geodesic dust. Observe that in the dissipative case the dust condition does not imply that the fluid is geodesic as can be easily seen from (\ref{2}), therefore the geodesic condition here is not redundant and is assumed, both,  for sake of simplicity and also in order to isolate the effects produced by dissipative phenomena alone.  It is worth mentioning that  a great deal of work has been done in order to find exact solutions describing radiating geodesic dust  configurations (e.g. see \cite{LTB}--\cite{36} and references therein)

Thus we have $P_r=P_{\perp}=0$ and $A=1$, in which case our equations (\ref{6}) and (\ref{7}) become
\begin{eqnarray}
\left({\cal E} - 4\pi \tilde  \mu \right)^{\dot{}}+\frac{3\dot R}{R}\left({\cal E}-4 \pi \tilde \mu \right)-\frac{12\pi \tilde q R'}{BR} =0,
\label{6dd}
\end{eqnarray}
and 
\begin{eqnarray}
\left({\cal E}- 4\pi \tilde \mu \right)^\prime=-\frac{3 R^\prime}{R}{\cal E }-\frac{12\pi \tilde q \dot R B}{R}.
\label{7dd}
\end{eqnarray}
From the equation above it follows that the quantity that now determines the existence of  energy--density inhomogeneities is:
\begin{equation}
\Psi\equiv{\cal E}+\frac{12\pi \int^r_0\tilde qB\dot R R^2dr}{R^3}.
\label{dis7}
\end{equation}

Indeed, from (\ref{7dd}) is not difficult to find that $\tilde\mu'=0 \leftrightarrow \Psi=0$.

Then, an equation for the evolution of $\Psi$ can be obtained from (\ref{6dd}), using (\ref{5b1}) and (\ref{1}) :
\begin{equation}
\dot \Psi-\frac{\dot \Omega}{R^3}=-4\pi \tilde \mu\sigma-\frac{4\pi \tilde q'}{B}+\frac{4\pi \tilde q R'}{BR}-\frac{3\Psi \dot R}{R},
\label{7disev}
\end{equation}
with $\Omega\equiv 12 \pi \int^r_0\tilde q B\dot R R^2 dr$. 

The general solution of (\ref{7disev}) is
\begin{equation}
\Psi=\frac{\int^t_0(\dot \Omega-4\pi \tilde \mu\sigma R^3-\frac{4\pi R^3\tilde q'}{B}+\frac{4\pi R^2\tilde q R'}{B})dt}{R^3},
\label{int7dis}
\end{equation}
where the influence of diferent factors on the evolution of $\Psi$ are clearly indicated.

We can further transform the expression above by noticing that in the  case under consideration (\ref{2}) can be formally integrated to obtain
\begin{equation}
\tilde q=\frac{\phi(r)}{B^2R^2},
\label{qdis7}
\end{equation}
where $\phi(r)$ is an arbitrary function satisfying $\phi(0)=0$.

Thus we see from (\ref{int7dis}) that starting from an homogeneous configuration  the appearance of density inhomogeneities will depend on two different factors; on the one hand the shear of the fluid and on the other dissipative terms. In order to isolate further the influence of these  latter terms, let us consider the  shear--free case.

If the shear is assumed to vanish, then it is a simple matter to see that  we may put $R=Br$, then using (\ref{qdis7}), we obtain from (\ref{int7dis})
\begin{widetext}
\begin{equation}
\Psi=\frac{\int^t_0\left[4\pi\int^r_0 \dot \Theta \phi(r)r dr-\frac{4\pi \phi(r)'r^3}{R^2} +\frac{8\pi \phi(r) r^2}{R^3}(\frac{5R'r}{2}-R)\right]dt}{R^3},
\label{int7disbis}
\end{equation}
\end{widetext}
from the above it is apparent that dissipation affect $\Psi$ through, both,  local and non--local tems.

Finally it is instructive  (always in the shear--free case) to bring out the role  of relaxational  effects in the evolution of $\Psi$. Thus we shall assume that we are in the purely diffusion approximation ($\epsilon=0$) implying $\tilde q=q$; $\tilde \mu=\mu$
Thus, from (\ref{2}) it follows
\begin{equation}
\dot q=-\frac{4q\Theta}{3},
\label{dif7}
\end{equation}
then combining the above equation with (\ref{ISE'bis}) we obtain
\begin{equation}
q=-\kappa \frac{rT'}{R(1-\frac{4\Theta \tau}{3})},
\label{tauq7}
\end{equation}
or using (\ref{qdis7})
\begin{equation}
\phi(r)=-\kappa \frac{R^3T'}{r(1-\frac{4\Theta \tau}{3})}.
\label{tauq7bis}
\end{equation}
Feeding back (\ref{tauq7bis}) into (\ref{int7disbis}) we can evaluate the influence of $\tau$ on the evolution of $\Psi$ for any specific configuration.
\section{CONCLUSIONS}
We have identified the different aspects of the fluid distribution which are responsible for the appearance of energy--density inhomogeneities and have found the evolution equation for  the variables representing those aspects. 

For dust and isotropic perfect fluids the relevant factor is the Weyl tensor, therefore the stability of the homogeneous energy--density  condition is equivalent to the stability of the conformal flatness condition.

However for locally anisotropic fluids, inhomogeneities are  also  related to the local anisotropy of pressure. The specific variable related  to the existence of energy--density inhomogeneities was found  as well as its evolution equation. That specific variable was identified as one of the structure scalars.

Dissipation emerges also as a source of energy--density inhomogeneities. In order to isolate the effect of dissipation from others factors we have considered the dust, geodesic case, and as a particular subcase we studied further the shear--free case.  The non--local contributions of dissipation to the formation of  inhomogeneities and the role  of relaxational effects are the most relevant results of this part of the study.

To conclude, the following remark is in order: The fact that the different variables considered here affect the formation of energy--density inhomogeneities might be intuitively clear, however the specific form in which these variables do so, is far from obvious. Indeed, we may consider a situation in which any of the above mentioned variables are present but are such that their combinations in  (\ref{ndd2}),  (\ref{anddios2}) or (\ref{int7disbis}) vanish. In this latter case the homogeneous energy--density condition will be  stable in spite of the fact that all (or some ) of the mentioned factors are present.

\end{document}